# Non-destructive study of collision cascade damage in self-ion irradiated tungsten using HR-EBSD and ECCI

Hongbing Yu[1*], Phani Karamched[2], Suchandrima Das[1], Junliang Liu[2], Kenichiro Mizohata[3], Felix Hofmann[1†]

(1) Department of Engineering Science, University of Oxford, Parks Road, Oxford, OX1 3PJ, UK

(2) Department of Materials, University of Oxford, Parks Road, Oxford, OX1 3PH, UK

(3) University of Helsinki, P.O. Box 64, 00560 Helsinki, Finland

[*] *hongbing.yu@eng.ox.ac.uk*       [†] *felix.hofmann@eng.ox.ac.uk*

**Abstract**: Understanding defect production and evolution under irradiation is a long-standing multi-scale problem. Conventionally, experimental examination of irradiation-induced defects (IIDs) has mainly relied on transmission electron microscopy (TEM), which offers high spatial resolution but only limited strain sensitivity (strains less than 0.1% are challenging to evaluate). TEM also requires very thin samples, making multi-scale characterisation and quantitative strain measurements difficult. Here, we explore the potential of using advanced techniques in the scanning electron microscope (SEM) to non-destructively probe irradiation damage at the surface of bulk materials. Electron channelling contrast imaging (ECCI) is used to image nano-scale irradiation-induced defects in 20 MeV self-ion irradiated tungsten, the main candidate material for fusion reactor armour. The results show an evolution of the damage microstructure from uniformly and randomly distributed nano-scale defects at 0.01 dpa (displacement per atom) to string structures extending over hundreds of nanometres at 1 dpa. Cross-correlation based high-resolution EBSD (HR-EBSD) is used to probe the lattice strain fields associated with IIDs. While there is little strain fluctuation at 0.01 dpa, significant heterogeneity in the lattice strains is observed at 0.1 dpa, increasing with dose until saturation at 0.32 dpa. The characteristic length scale of strain fluctuations is ~500 nm. Together, ECCI and HR-EBSD reveal a transition from a structure where defects are disordered to a structure with long-range order driven by elastic interactions between pre-existing defects and new cascade damage.

1. Introduction

Understanding irradiation-induced defect (IID) structures, their evolution and the consequent dynamics of changes in material properties is a complex, multi-length- (sub-atomic to macro-scale) and time-scale (femtosecond to tens of years) problem [1–3]. There are two established pathways to improving this understanding: Computational multi-scale modelling, where simulations are used to link physics at different scales, [1,2,4,5], and multi-technique experiments that give insight into IID production and evolution and serve to validate the predictions of the computational models [6–13].

Both paths have certain advantages and limitations. In computational models, the constraint on the proportionality between length and time scales (e.g. atomic-scale modelling cannot extend beyond few nanoseconds), can compromise capturing of real-time microstructural evolution. Concerning experimental techniques, it is challenging to investigate extremes of time and length, i.e. very fast and very slow processes, or nano-scale spatial resolution. Transmission electron microscopy (TEM) is the most widely used tool for direct examination of defect structures in irradiated materials [14]. Time scales spanning from 5 ms to a few hours can be investigated through *in situ* TEM observation [15], while length scale of 2 nm to a few microns can be probed with TEM diffraction techniques [16]. High energy X-ray diffraction can extend the examined length scale to μm-mm, but only produces statistical information [9,17]. Recent advances in coherent X-ray diffraction imaging (CDI) have enabled the direct imaging of dislocation loops with a spatial resolution of 20 nm [18,19]. However, the size of the sample used for CDI is limited by the coherence length of the X-ray beam, which is ~ 1 μm.

Previous TEM studies of heavy ion damage in body-centred cubic (bcc) materials have examined defect production, interactions between IIDs and the evolution of damage structures [7,10,16,20–24]. Though detailed defect production and evolution mechanisms may vary from material to material [16,22,25], the general trend is very similar for bcc materials. Generally, it has been found that the defect number density varies exponentially with irradiation dose at lower doses, before reaching a saturation level beyond a certain threshold dose (the threshold varies depending on the material). At low doses, IIDs in TEM micrographs have predominantly dot-like morphology [23,26,27]. At higher doses, IIDs grow to loop-like appearance with long-range spatial ordering and eventually forms string structures or complex network structures [23,26,27]. Dose-dependent microstructural evolution and/or the nature and relative proportion of dislocation loops Burgers vectors (e.g. ½ <111> or <100>) are strongly influenced by parameters such as temperature, grain orientation and the presence of alloying elements [16,22,27]. A key reason for the strong dependence on the above-mentioned parameters is the elastic strain field of IIDs.

Each IID, no matter how large or small, gives rise to an elastic strain field. This affects defect structure evolution at all dose levels. For instance, elastic trapping of dislocation loops within a cascade can prevent large prismatic loops from escaping to nearby surfaces [28]. On the other hand surface image forces can lead to substantial loss of ½ <111> dislocation loops near-surfaces [10,29] and thus result in excessive <100> loop populations in thin TEM foils, especially in Fe [30]. This illustrates that the damage structure observed in thin foils by TEM can be significantly altered by surface effects. Mutual elastic interactions of loops also cause loop reactions, coalescence and self-organisation of loops [16,20,23]. These interactions can eventually result in the formation of long-range raft structures of loops with collinear Burgers vector [26]. To make a complete, physically-based prediction of microstructural evolution in irradiated metals, the interplay between strain evolution and evolution of IIDs must be accounted for.

The strains field induced by IIDs can be non-invasively probed, with depth resolution, using synchrotron X-ray micro-Laue diffraction [12,31] or Bragg coherent X-ray diffraction imaging (BCDI) [18,19,32]. However, access to synchrotron sources is limited and slow. Alternative techniques of 2D strain-mapping are TEM based approaches such as geometric phase analysis (GPA) [33] and nano-beam diffraction [34], though these have limited strain sensitivity (on the order of ~ $10^{-3}$) [35,36]. Focused-ion beam (FIB) milling or other thinning approaches are generally used to prepare thin samples for TEM investigation from more macroscopic irradiated material volumes [37,38]. This thinning process partially relieves residual stresses that were present in the material following irradiation, making it challenging to directly correlate the microstructure of the observed volume to its properties (a compromised post-facto method commonly used). Further, FIB milling itself can introduce significant irradiation damage in thin foils [19], which causes confusion when interpreting the observed microstructure [39–41] and can give rise to large errors in strain measurement [18,19]. As such, TEM-based techniques have been rarely used to monitor strain evolution in irradiated materials.

There is an urgent need for the development of non-destructive, multi-scale, easy-to-access, high-sensitivity strain measurement techniques. Two emerging SEM techniques, electron channelling contrast imaging (ECCI) and high angular resolution EBSD (HR-EBSD) [42–48], have great potential for characterization of IIDs. ECCI utilises the strong dependence of the backscattering intensity on the diffraction conditions [46,48]. At the channelling condition, the back-scattered intensity is very low. Even slight local lattice distortions, due to dislocations, change the diffraction condition and cause a strong backscattering of electrons, thus forming a sharp contrast near dislocations or lattice distortions [46,48]. As such the contrast is qualitatively rather similar to weak-beam dark field (WBDF) TEM micrographs.

HR-EBSD uses cross-correlation to determine small shifts of specific zone axes relative to the diffraction pattern measured at a reference point. Since these small shifts are linked to distortion in the lattice, the full deviatoric strain tensor and lattice rotation vector can be probed by HR-EBSD, with a sensitivity of ~$2\times10^{-4}$ [43,45,46]. HR-EBSD has been used to characterize the spatial distribution of geometric necessary dislocations (GNDs) and elastic strain tensor around specific microstructural features of interest, such as nano-indents [49], grain boundaries [50], phase interface and cracks [51], and slip bands [52]. Improvements in spatial resolution also enabled measurement of the lattice strains and rotations caused by a single dislocation [35]. Despite promising advances, ECCI and HR-EBSD have not been widely adopted to characterise IIDs, except in a recent work demonstrating reliable imaging of irradiation-induced nanoscale voids by back-scatter electron imaging in SEM [53].

Here we present a prototypical non-destructive study of the evolution of IIDs and their associated strain fields as a function of self-ion irradiation dose, using a combination of ECCI and HR-EBSD in a bulk sample. Non-destructive here refers to the fact that no further sample preparation or cutting is required provided well-polished sample surfaces were prepared before irradiation and are maintained throughout the irradiation processes. We concentrate on pure tungsten, which is an attractive bcc model material, as well as the current front runner material choice for plasma-facing armour in future fusion reactors [2,54]. ECCI is used to image defect structures as a function of irradiation dose, while HR-EBSD is used to probe the full deviatoric strain tensor. The combination of ECCI and HR-EBSD measurement reveals an evolution from a structure where defects are randomly distributed at low doses to long-range ordering at higher dose, driven by elastic interactions between pre-existing defects and new cascade damage. This provides a new window for gaining insight into IID evolution under collision cascade damage, and the role played by elastic interactions.

## 2. Materials and experiments

High-purity polycrystalline tungsten (>99.97 wt. %, purchased from Plansee) was used for this study. The as-received 1 mm thick plate was cut into 10 mm × 10 mm × 1 mm samples. They were annealed at 1500 °C for 24 h in a vacuum furnace (~$10^{-5}$ mbar), resulting in a fully recrystallized microstructure as shown in Fig. 1 (a). After heat treatment, the samples were ground with abrasive papers from 600 grade down to 1200 grade, followed by diamond paste polishing from 6 μm to 0.5 μm grade. A final electropolishing step, using an electrolyte of 1% NaOH in deionized water, at a voltage of 8 V at room temperature (RT) for 1 to 2 minutes, produced a defect-free mirror surface finish. The density of geometric necessary dislocation (GND) in the as-polished samples is on the order of $10^{13}\ m^{-2}$, measured by HR-EBSD using a step size of 55 nm.

Self-ion irradiation was used to mimic the damage produced in tungsten during neutron irradiation. An ion energy of 20 MeV was chosen as this produces a ~ 2 μm thick damage layer, which allows further micro-mechanical study (e.g. nano-indentation) and thermal transport measurements (e.g. using transient grating spectroscopy [55]) of the damaged material. The samples were irradiated at RT to different damage levels from 0.01 dpa to 1 dpa. The anticipated displacement damage was estimated using the of Stopping and Range of Ions in Matter (SRIM) code [56] (K-P calculation, threshold energy of 68 eV [12]). The calculated damage profile is shown in Fig. 1(b). Based on the calculation, $2.53 \times 10^{14}$ ions/cm$^2$ are required to reach 1 dpa at the peak position. Ion irradiation was carried out at the Helsinki Accelerator Laboratory, with 20 MeV $^{184}$W$^{5+}$ ions, with a target current density of ~ 25 − 40 nA/cm$^2$, corresponding to a dose rate of $1.2-1.6\times10^{-4}$ dpa/s. Raster scanning was utilised to obtain a spatially uniform implantation profile across the sample surface. Details of the implantation can be found elsewhere [57,58].

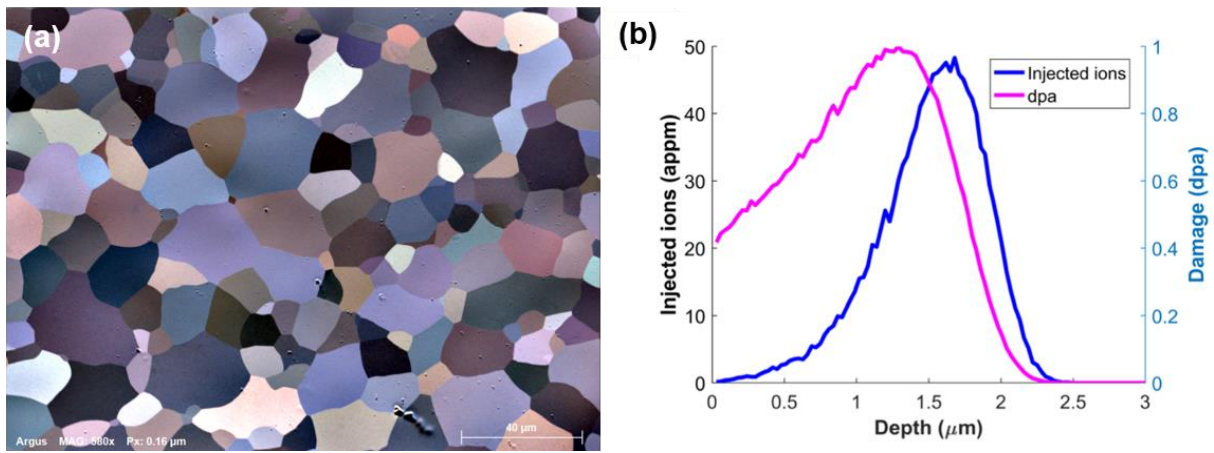

Fig. 1 (a) Argus image showing the recrystallized microstructure after heat treatment at 1500 °C for 24 h. (b) Damage profile and the injected ion distribution along the depth direction in the 1 dpa sample.

ECCI was performed on a Zeiss Crossbeam 540 SEM equipped with a backscattered electron (BSE) detector. The most critical aspect of imaging lattice defects using ECCI is to correctly set up the two-beam condition, where electrons can channel between lattice planes, with little back-scatter [48]. To achieve this condition, the sample was tilted to ~ 7°, and the contrast of the region of interest monitored with the BSE detector whilst rotating the sample about the surface normal. A good two-beam condition is achieved when a sharp change in the contrast is observed due to a small rotation (essentially this corresponds to the edge of a Kikuchi band [59,60]). The diffraction vector (**g** vector) of the ECCI images was determined by EBSD. The voltage and current used for this study are 30 keV and 10 nA, respectively.

A Zeiss Merlin SEM with a Bruker eFlash detector was used to carry out HR-EBSD measurements (20 kV, 15 nA). The normal EBSD set up, with the sample tilted at 70° to the electron beam, was used. The diffraction pattern at each point was recorded at a resolution of 800 ×600 pixels. Working distance and the sample to detector distance are both 18 mm. A step size of 55 nm was used for all EBSD maps. Cross-correlation was used to quantify small shifts of specific zone axes at each point relative to a reference point [45]. The full deviatoric strain tensor and lattice rotation tensor can be computed if shifts of 4 or more zone axes are known. Since the diffraction pattern is insensitive to lattice dilation, the strain we probe here represents the deviatoric part. The cross-correlation analysis of the Kikuchi patterns was done using the XEBSD code described by Britton & Wilkinson [61,62].

## 3. Results

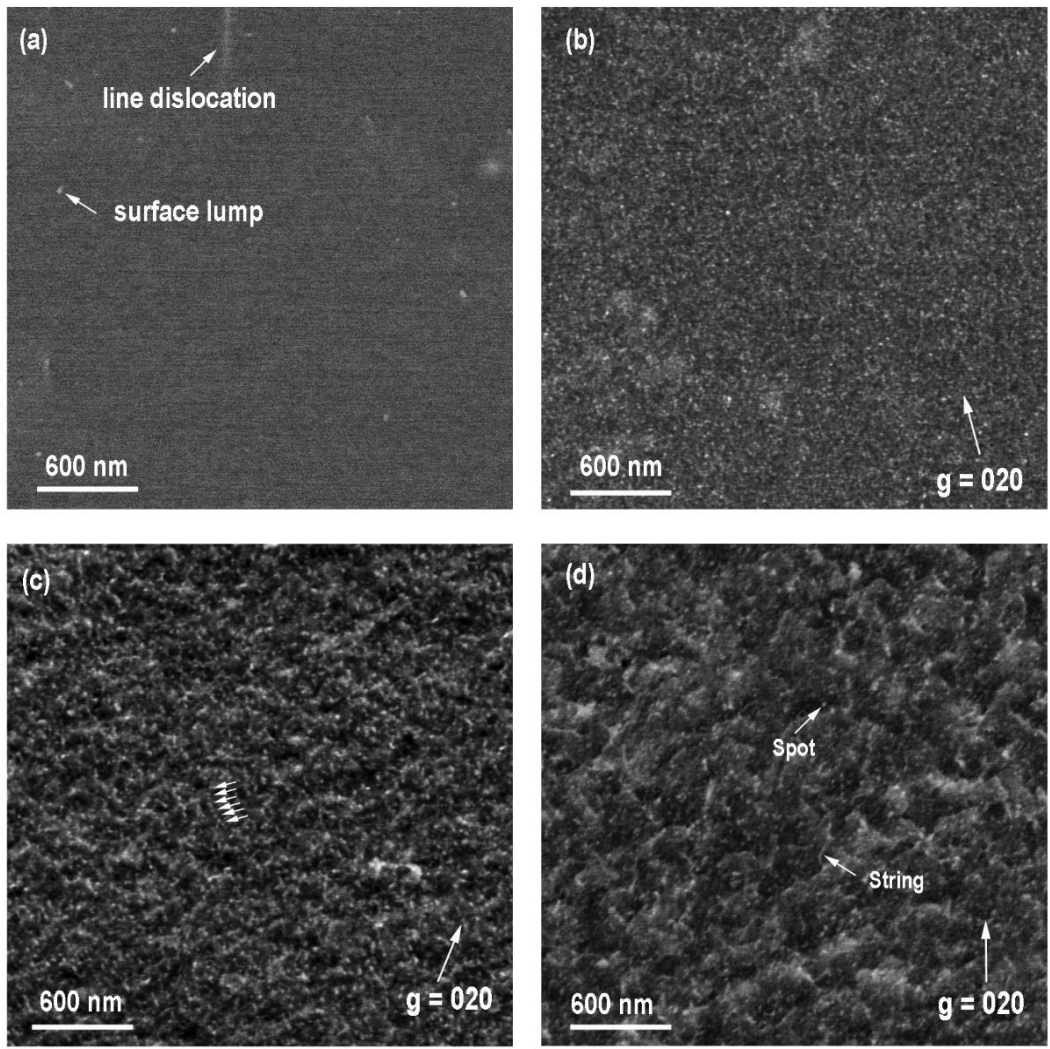

Fig. 2 ECC1 images of pure tungsten samples irradiated by 20 MeV self-ions to different dose levels. (a) 0 dpa, (b) 0.01 dpa, (c) 0.1 dpa, (d) 1 dpa. The images were all taken on grains with {100} surface orientation.

ECCI images from an unimplanted reference sample and samples exposed to damage levels of 0.01 dpa, 0.1 dpa and 1 dpa are shown in Fig. 2. The pixel size of ECCI images is 4.6 nm for the un-irradiated sample or 2.2 nm for the irradiated samples. Several grains in the unirradiated sample were checked with multiple g vectors. There are few lattice defects. Fig. 2a shows the ECCI image of the un-irradiated sample, a dark background decorated with a bright line and a few bright spots. The bright line is typical contrast from a line dislocation [42]; while the bright spots are mainly from surface imperfections left by electropolishing (see supplementary Fig. S1 for a comparison between ECCI and secondary electron images).

In all irradiated samples, ECCI images were taken under a two-beam condition with 020 **g** vector. The 0.01 dpa sample shows dense and uniformly distributed small bright spots (Fig. 2b). These bright spots are contrast arising from nano-scale irradiation-induced dislocation loops. The local lattice distortions associated with these defects enhance the back-scattering of electrons, thus giving bright contrast. At 0.1 dpa, the damage structure is also dominated by dense nano-scale spotty dislocation loops. However, the apparent size of the loops is larger than in the 0.01 dpa sample. To allow a quantitative comparison between 0.01 dpa and 0.1 dpa, the number density and the image size of the dislocation loops were measured. This was done using the automated counting algorithm described by Mason [63]. For a detailed description, please refer to supplementary Fig. S2. Considering the limitation on the spatial resolution of the SEM (~ 3 nm) when applied to bulk material [48], only loops with a diameter larger than 4 nm were considered. Averaging over three $3 \times 3$ μm$^2$ ECCI images for each dose, there are 4643 (±4%) loops and 3355 (±7%) loops in the 0.01 dpa sample and 0.1 dpa sample, respectively. This corresponds to an averaging areal number density of $5.1 \times 10^{14}$ m$^{-2}$ and $3.7 \times 10^{14}$ m$^{-2}$, respectively, i.e. there is a decrease in the areal number density of dislocation loops from 0.01 to 0.1 dpa. At the same time, there is a significant increase in dislocation loop size from 0.01 to 0.1 dpa, visible as a shift to the right of the size distribution peak in Fig. 3. If we consider the total number of point defects contained in these dislocation loops [64], the 0.1 dpa sample has ~15% more point defects stored in loops. Besides, some clustering of loops can also be observed in the 0.1 dpa sample (Fig. 2(c)).

The 1 dpa sample shows a rather different damage structure, consisting of a mixture of string structures, as well as spotty dislocation loops. The string structures must have formed as a result of the interaction of new collision-cascade-induced defects with pre-existing dislocation loops as the damage level is increased beyond 0.1 dpa. The quantitative information for the 1 dpa sample is hard to retrieve using the counting algorithm of Mason [63], because of the existence of complex structures. Qualitatively, the 1 dpa sample image suggests that the number density of spotty loops decreases significantly once string structures are formed. However, the spotty loops do not disappear altogether. Those that are visible must have been either formed by recent collision cascades, or they may be residual loops that were formed at a lower dose and have been retained.

It is interesting to consider the length scale of the clustering and string structures, which can be visualised by the rotational average of the 2D fast Fourier transform (FFT) of the ECCI images [65]. For a detailed description, please refer to supplementary Fig. S3. Fig. 4 shows the rotational average of the 2D FFT of the 4 ECCI images in Fig. 2. The x-axis represents the frequency domain, f, which is equivalent to length in reciprocal space. That is, 1/f corresponds to length in real space. For the un-implanted sample, the ECCI image is dominated by background signals. The magnitude of the rotational average of the FFT decreases rapidly with the increase of the frequency up to $3\times10^{-3}$ nm$^{-1}$, and then saturates. Though a monotonic decrease with frequency is also seen in the profile of the 0.01dpa sample, the shape is distinctly different from the un-implanted case. After an initial decrease, a steady-state with reduced gradient is seen in the interval between $1\times10^{-3}$ and $2\times10^{-2}$ nm$^{-1}$. This is followed by another more rapid decrease and final saturation at $7\times10^{-2}$ nm$^{-1}$. The profile of the 0.01 dpa sample has a higher magnitude at the steady-state stage ($1\times10^{-3}$ to $2\times10^{-2}$ nm$^{-1}$) than the unimplanted one, but lower magnitude at the final saturation stage. This is due to the presence of a dense population of small dislocation loops.

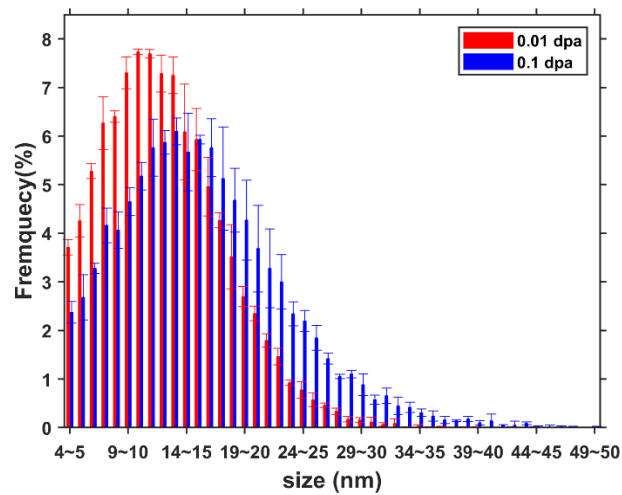

Fig. 3 Size distribution of dislocation loops in 0.01 dpa and 0.1 dpa samples.

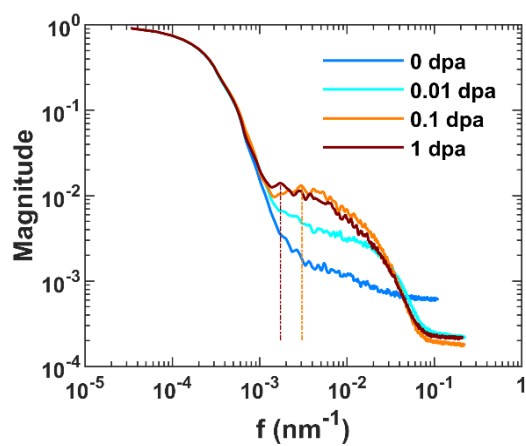

Fig. 4 Rotational average of the 2D FFT of the ECCI images plotted on a log-log scale. The x-axis is the frequency domain (f), which is equivalent to the length in reciprocal space.

The 2D FFT profile evolves significantly with further increasing damage dose. At damage levels of 0.1 dpa and 1 dpa, the profile no longer decreases monotonically, but a peak appears in the interval between $1\times10^{-3}$ and $4\times10^{-3}$ nm$^{-1}$. There is also an increase in magnitude in the interval between $1\times10^{-3}$ and $2\times10^{-2}$ nm$^{-1}$ in the 0.1 dpa profile compared to the 0.01 dpa profile. There is no significant further evolution of the FFT profile from 0.1 dpa to 1 dpa.

It is interesting that apart from the changes in the interval between $1\times10^{-3}$ and $2\times10^{-2}$ nm$^{-1}$, the profiles of the three implanted samples agree closely for the rest of the frequency domain. This suggests that the FFT profile in this interval must be directly linked to the spatial frequencies associated with the defect structures. The steady-state decrease for 0.01 dpa corresponds to a uniform distribution population of small dislocation loops. The peak for the 0.1 dpa sample at f = $3\times10^{-3}$ nm$^{-1}$ reflects the clustering of dislocation loops and suggests an associated characteristic length scale of ~ 330 nm. The corresponding characteristic length scale of the complex structures in the 1 dpa sample is slightly larger ~ 500 nm.

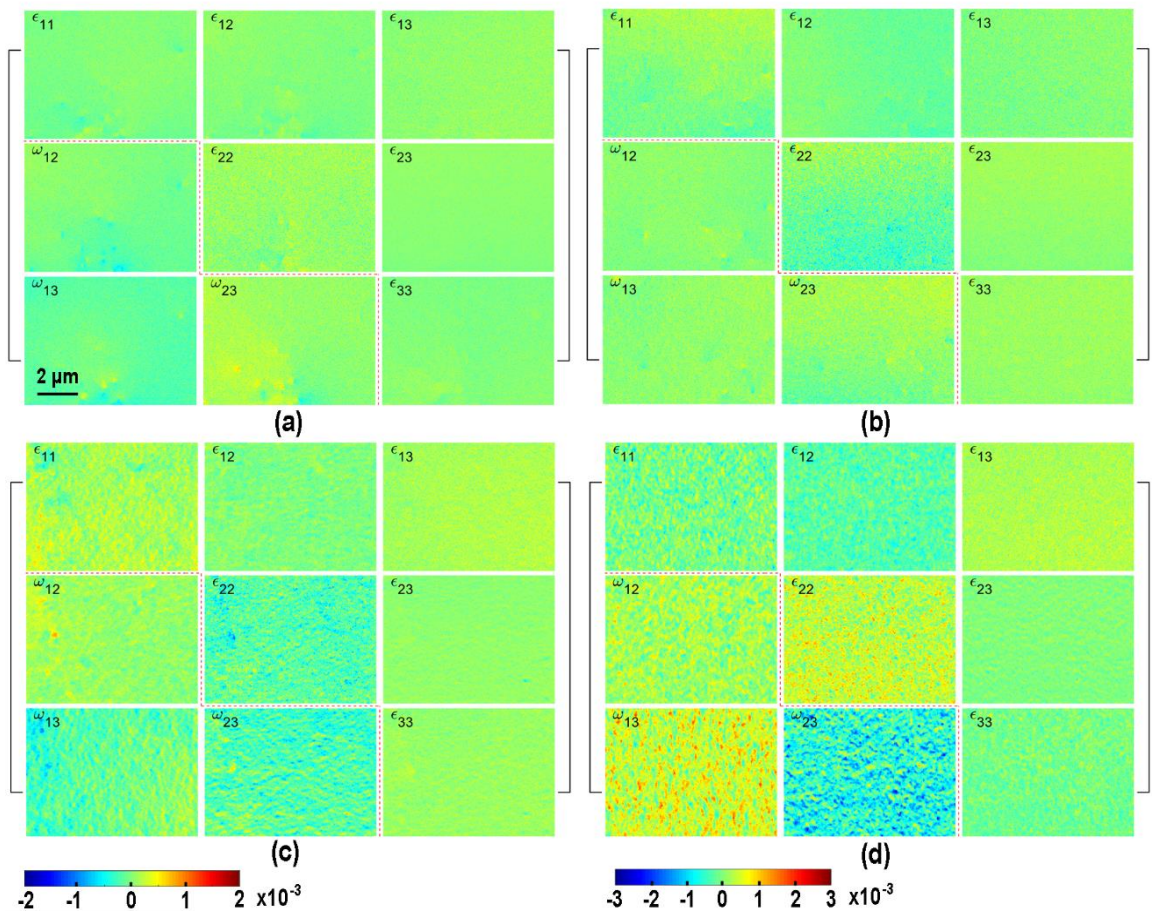

Fig. 5 The evolution of the lattice strains (ε) and lattice rotations (ω) as a function of irradiation damage dose. The colour scale of (a-c) is [-$2\times10^{-3}$, $2\times10^{-3}$], but (d) is [-$3\times10^{-3}$, $3\times10^{-3}$]. (a) 0 dpa, (b) 0.01 dpa, (c) 0.1 dpa, (d) 1 dpa. The scale bar in (a) applies to all parts of this figure.

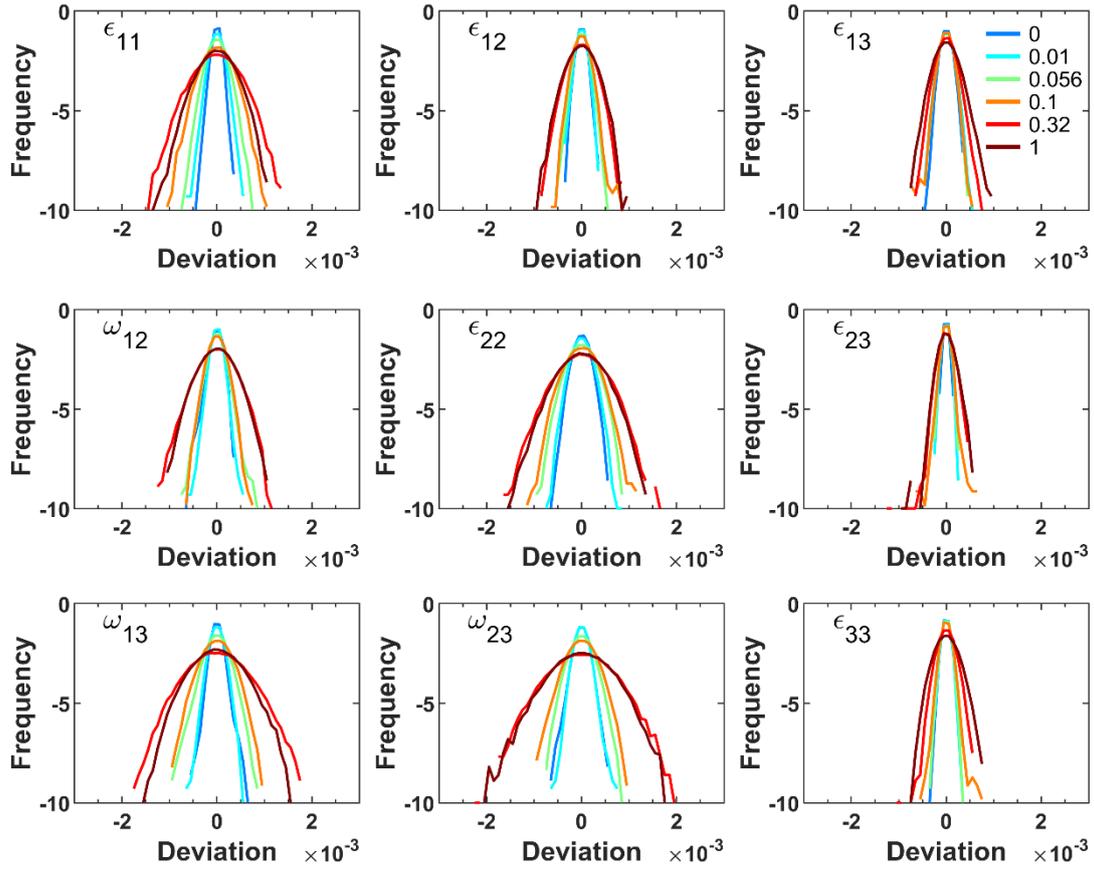

Fig. 6 Histogram of deviation from the mean for all lattice distortion components. For lattice rotation components the x-axis is in radians. The colour-coding in the $\epsilon_{13}$ component refers to the sample dose in dpa, and applies to all components.

The lattice strain field evolution associated with the defect structures is shown in Fig. 5. The full deviatoric lattice strain tensor and lattice rotations from 0 dpa up to 1 dpa are plotted. The size of the maps is 9.5× 7.2 μm, and a stepsize of ~55 nm was used. It should be pointed out that the deviatoric lattice strains and rotations for the implanted samples are relative rather than the absolute values since no strain-free reference is accessible in these samples. As such these maps reflect the spatial variation or fluctuation of the strain and rotation fields. The unirradiated sample was well annealed and shows few residual defects in ECCI (Fig. 2 (a)), and hence little residual lattice strain. The variation in the 6 lattice strain components and 3 lattice rotations is very small (Fig. 5a), as expected. A few hot spots in the lower half of the strain and rotation maps may correspond to residual dislocations rather than surface imperfections (see supplementary Fig. S4). The variation of lattice distortions in the 0.01 dpa sample is as small as in the unimplanted sample, even though a high density of nano-sized IID is observed by ECCI. The quantitative evolution of the magnitude of the fluctuations with irradiation dose is shown in Fig. 6, as histograms of the deviation from the average for all strain and rotation components. It should be noted that additional doses (0.056 dpa and 0.32 dpa) are included in Fig. 6.

The full width half maximum (FWHM) of these histograms represents a quantitative measure of the magnitude of the fluctuation. Little difference is observed between 0 dpa and 0.01 dpa in all the lattice strain components. The absence of fluctuation in lattice strains and lattice rotation at the length scale of 50 nm suggests that there is no long-range ordering of dislocation loops with the same Burgers vectors. This is consistent with ECCI observation (Fig. 2(a) and Fig. 4) showing the random distribution of nano-scale dislocation loops in the 0.01 dpa sample.

Long-range fluctuation begins to emerge in the lattice strain components, especially in the normal strains, and lattice rotations at 0.1 dpa (Fig. 5c). These fluctuations become more prominent with increasing damage dose, as seen in Fig. 5d. This is quantitatively presented in Fig. 6. A clear trend is seen that the magnitude of the fluctuations increases with the damage level from 0.01 dpa onwards until 0.32 dpa in all the components expect $\varepsilon_{23}$. The change of the magnitude of fluctuations with damage dose is monotonic and gradual except in components $\varepsilon_{12}$ and $\omega_{12}$, where an abrupt change between 0.1 dpa and 0.32 dpa is seen. After 0.32 dpa, a saturation of the fluctuations is observed in all the components.

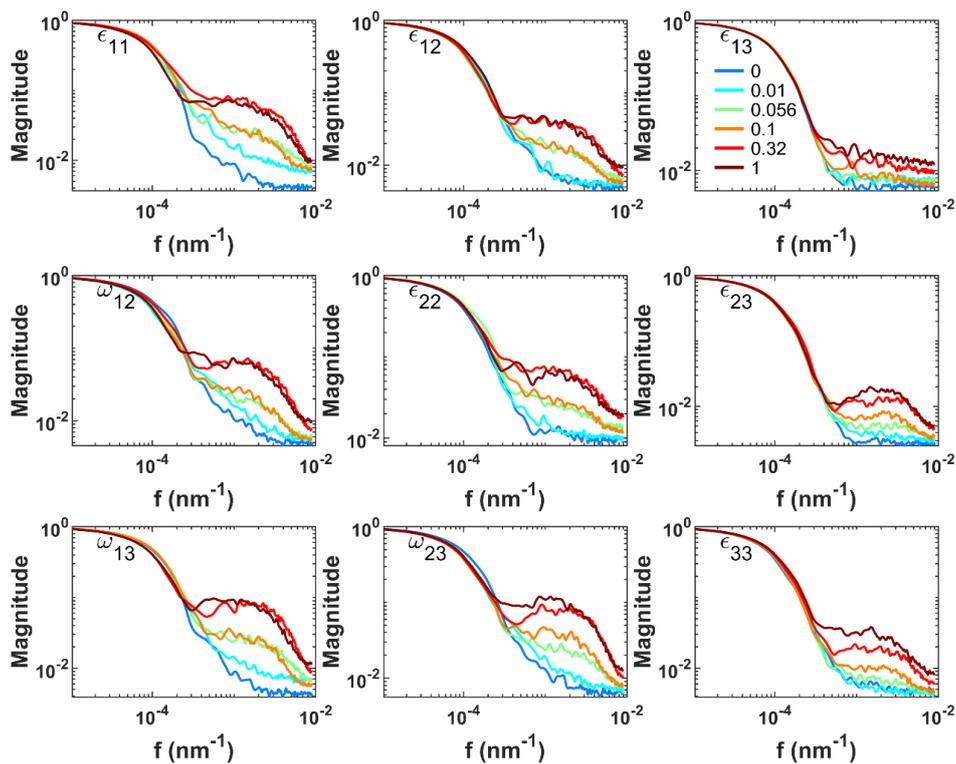

Fig. 7 Rotational average of the 2D FFT of the lattice strain and lattice rotation maps measured by HR-EBSD. The colour-coding in the $\epsilon_{13}$ component refers to the corresponding damage levels in dpa and the same colour-coding is used in all parts of this figure.

To reveal the length-scale of the strain fluctuation, the "2D rotational average of the FFT" analysis was performed on the strain and rotation maps, as shown in Fig. 7. With the increase of dose, the most significant changes occur in the interval between $3\times10^{-4}$ nm$^{-1}$ and $10^{-2}$ nm$^{-1}$. Here we focus on this interval. The evolution of the FFT profile with damage dose can be divided into 3 stages. The first stage corresponds to 0.01dpa and below, where no significant difference is observed compared to the unimplanted sample (see Fig.7 except the $\varepsilon_{11}$ component). The second stage is at dose levels between 0.056 dpa and 0.32 dpa, where a transition behaviour is seen. At this stage, the magnitude of the FFT increases significantly with dose, especially in the interval between $5\times10^{-4}$ nm$^{-1}$ and $5\times10^{-3}$ nm$^{-1}$. Gradually, peaks start to appear in some components with the increasing dose, for example in the $\varepsilon_{23}$, $\varepsilon_{33}$, $\omega_{12}$, and $\omega_{23}$ components for the 0.1 dpa sample. Eventually, peaks appear in all components except the $\varepsilon_{13}$ component. Finally, for damage levels above 0.32 dpa, the FFT profile does not change significantly with dose, suggesting a saturation state. The peaks fall into the frequency range from $1\times10^{-3}$ nm$^{-1}$ to $2\times10^{-3}$ nm$^{-1}$. That is, the fluctuation in the lattice strain fields and lattice rotation fields have a characteristic wavelength on the order of ~500 nm. This agrees closely with the length scale of the ordering we found in ECCI.

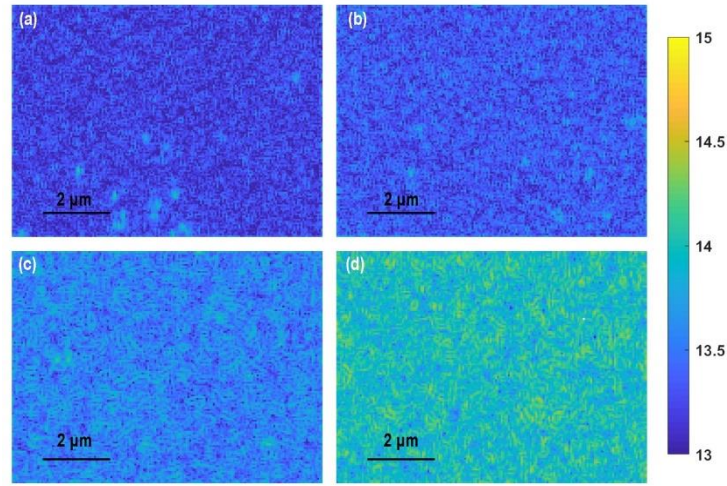

Fig. 8 GND maps of samples irradiated to different dose levels. (a) 0 dpa, (b) 0.01 dpa, (c) 0.1 dpa, (d) 1 dpa. The colour map shows GND density in m$^{-2}$ on a log scale with base 10.

The density of geometric necessary dislocations (GND) can be computed from HR-EBSD measurements [66,67]. Since a 55 nm step size was used, GND density here represent the geometrically necessary dislocation population at a 55 nm length scale. This is important since the measured GND density shows scale dependence. The smaller the step size, a higher fraction of the total dislocation density is geometrically necessary [68]. Fig. 8 shows the distribution of GND density in the four samples. The average GND density as a function of dose level is plotted in Fig. 9. Little increase in the GND density is observed from 0 dpa to 0.01 dpa though a lot of IIDs have been induced, as evident from ECCI (Fig. 2 (a)). After 0.01dpa, a dramatic increase in GND density with irradiation dose is seen, even though there is only a small increase in the total number of point defects stored in dislocation loops at this stage. GND density saturates at 0.32 dpa, in line with the evolution of the strain fluctuation.

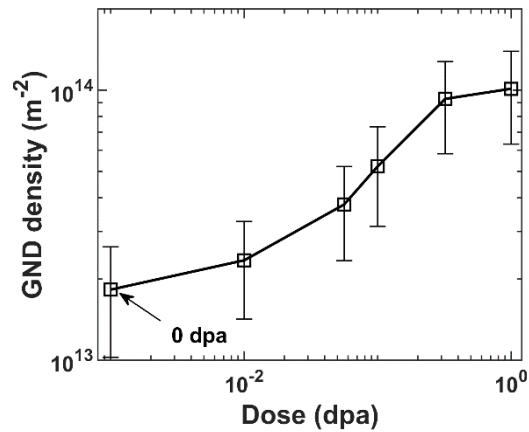

Fig. 9 Averaged GND density plotted as a function of damage level.

## 4. Discussion
### 4.1. Non-destructive study of the damage structure
#### 4.1.1. ECCI

The evolution of the damage structures observed non-destructively on the surface using ECCI agrees quite well with TEM observations of FIB lift-out cross-section samples of tungsten implanted with 20 MeV self-ions at room temperature [41]. Three stages of evolution: uniform and random distribution of nanoscaled loops at 0.01dpa, coarsening and clustering of loops at 0.1 dpa, and coexistence of dot-like loops and complex string structures at ~ 1 dpa, were observed from both techniques. However, since the FIB induced damage during sample preparation is hard to remove, this can lead to misinterpretation of the observed structures [39,41]. For instance, in [41] the spotty loops observed by TEM at ~ 1 dpa in 20 MeV self-ion implanted tungsten were excluded from the ion implantation induced damage, and attributed to FIB damage. However, our ECCI observations show that there are still a considerable number of dot-like loops even after the formation of string structures at ~ 1 dpa. The average defect size at 0.01 dpa was assessed as ~5 nm from TEM bright-field images of 20MeV self-ion implanted tungsten [41]. This agrees quite well with the average defect size of 4.6 nm, reported from TEM weak-beam dark field (WBDF) images of 150 keV self-ion irradiated tungsten [16]. The defect image size measured from ECCI is significantly larger than that determined from TEM images. Two factors may be responsible for this: Firstly, the spatial resolution of SEM on bulk materials is much lower than TEM on a thin foil sample.  For instance, at 0.01 dpa, small defects less than 4 nm dominate the size distribution profile [6], which ECCI is probably not sufficiently sensitive to pick up, resulting in a larger average size of defects. Secondly, the relationship between the defect image size and real defect size in ECCI has not been well studied [48].  In terms of the defect density, loop density measure from ECCI is about half of that determined with TEM for 150 keV self-ion implanted tungsten at 0.01 dpa [16]. Considering that for deformation-induced line dislocations, the dislocation density measured from two-beam ECCI is about half of that determined by TEM bright-field [59], the loop density evaluated from ECCI is in an accepted range. To sum up, while absolute quantification of defect number density and defect size with ECCI is challenging, it provides an attractive tool for studying trends and looking at defect structure evolution as a function of irradiation damage. It avoids the issue of FIB damage and is much faster than TEM characterisation, potentially enabling more rapid throughput non-destructive characterization.

### 4.1.2. HR-EBSD

It is interesting that HR-EBSD, despite its spatial resolution of tens of nanometers [69], can be successfully applied to the characterization of nano-sized IIDs. The 0.01 dpa sample has a dense, nano-scale IIDs population. Yet, no significant fluctuations in the lattice strains due to these defects were observed. Each dislocation loop, though small, gives rise to a local strain field. The lack of strain fluctuation at 0.01 dpa indicates that the resultant strain field, induced by the dislocation loops within the EBSD interaction volume, is uniform. Based on TEM observation, both $1/2<111>$ and $<100>$ dislocation loops can form in pure tungsten under collision cascade damage [7,16]. As such 14 different loop variants may be present in tungsten. The relatively uniform distribution of lattice strain suggests a random distribution of dislocation loops with random assignment of Burgers vector, as shown schematically in Fig. 10a. In this case, each dislocation loop within or surrounding the EBSD interaction volume distorts the lattice in a localized region. However, a convolution of this strain distribution with the EBSD interaction volume results in an averaging out of nano-scale local fluctuations, thus giving little average strain variation at the HR-EBSD spatial resolution of 55 nm.

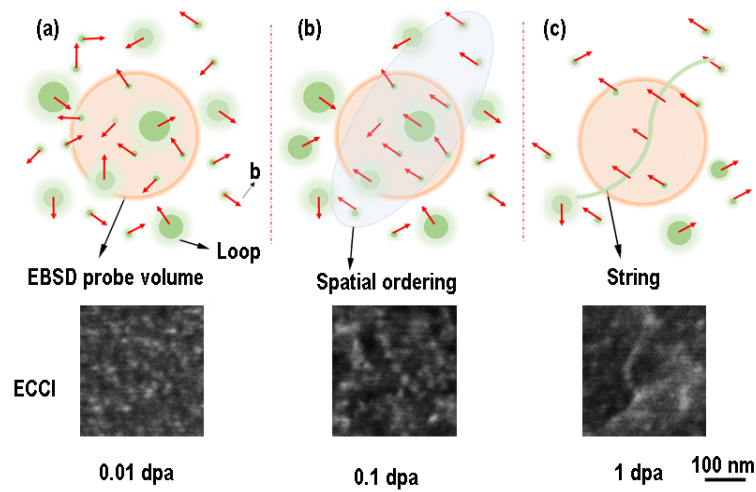

Fig. 10 Schematic and the corresponding ECCI images showing the evolution of the defect structures from randomly distributed dot-like loops to long-range, spatially-ordered structures. The orange circle represents the EBSD interaction volume. The red arrows represent Burgers vector, and the same scale bar applies to all ECCI images.

At 0.1 dpa, defects remain dot-like in morphology, though their average apparent size has increased to 16 nm from 13 nm at 0.01 dpa (see Fig.1 and 2). This means that there should only be a small increase in the strain field associated with individual dislocation loops. However, there are noticeably larger fluctuations in the lattice strain and rotation fields (Fig. 5). The length scale associated with these fluctuations is ~ 500 nm. This long-range strain fluctuation is in good agreement with the clustering of loops observed by ECCI at 0.1 dpa (Fig. 4). Some loops within the cluster are aligned into string configurations as shown in Fig. 10 (b). This arrangement suggests that they have collinear Burgers vectors (see Section 4.2) and as such displace the lattice collectively, giving rise to the longer-range fluctuations captured by HR-EBSD.

TEM has been widely used to characterize the evolution of irradiation damage. Nevertheless, due to the limit field of view and sample bending which causes significant contrast changes, the transition stage from randomly distributed dot-like loops to complex string structure has not been well captured [7,16]. Evolution at this stage is very important for the understanding of the resulting string structures that can substantially modify the behaviour of materials exposed to high doses [70]. By combining ECCI and HR-ESBD the transition in the microstructural evolution of IIDs can be revealed, as well as the transition in strain field and rotation field from short-range, localized strain fields (too small to be probed by HR-EBSD) to long-range fluctuation.

The evolution of GND density with damage dose deserves more discussion. The GND density saturates at 0.32 dpa, much higher than the dose at which the defect number density saturates (between 0.01 and 0.1 dpa). Saturation of GND density signifies that a steady-state long-range ordering of dislocation loops has been reached. This can be very useful information. The prediction of the mechanical response of the irradiated material relies on the prior knowledge of the loop density and loop size [57,71]. Again TEM observation has been the main source of quantitative information. Once the complex structure forms, quantitative analysis of defects seen in TEM images can be very difficult [72]. This is because the formation of complex structures makes it impossible to define the size and determine the number of dislocation loops [73]. At 1 dpa, ECCI reveals the coexistence of dot-like loops and complex string-like structure, while, the lattice distortion field shows long-range ordering. This suggests that the strain fields of the dotty loops are most likely aligned with the surrounding strain fields of the string structures. Though GND density was measured at a spatial resolution insufficient to probe every single dotty loop, it could be potentially used as a quantitative parameter for mechanical simulations once the long-range ordering forms.

### 4.2. Evolution of defects structures and configuration

Previous studies have demonstrated that there is a strong elastic interaction between loops produced by primary displacement cascades at low dose [6,28]. This elastic interaction can significantly affect the evolution of defects structures. For instance, it can bring loops together to form an elastic trap, preventing the escape of loops to the surface [28]. Though strong, elastic interactions for small loops are relatively short-range, limited to tens of nm. Previous studies did not show any evidence suggesting long-range ordering of dislocation loops at low doses [6,7,16]. This is in good agreement with the random loop distribution seen in ECCI (Fig. 2(b) and Fig. 10(a)) and the lack of strain fluctuations in the HR-EBSD measurement (Fig. 6 and 7) of the 0.01 dpa sample.

The complex string structure, i.e. string-like clusters of loops, observed in the room-temperature-implanted 1 dpa sample is interesting since no such structure has been observed when lower ion energies were applied. Previous TEM studies by Yi [7,16,20] showed that 150 keV or 2 MeV self-ion implantation did not produce complex string or web structures in <100> oriented grains at temperatures below 500 °C up to 0.8 dpa. Alloying elements such as Ta and Re can further prevent the formation of network structures in <100> grains at even higher temperatures [16]. Surprisingly, the observation of complex string structures has been reported to depend on grain orientation [7,16]. Complex string structures form more readily in <110> and <111> grains, at temperatures as low as 300 °C and 0.4 dpa. This orientation dependence of damage structures has been explained in terms of the traction effect of nearby surface on defects in TEM samples [7]. It was postulated that dislocation loops of different ½<111> variants have the same chance to escaping to the surface in <100> grains, but that there is a preferential loss of certain variants in non-<100> grains [7,16]. The resulting preferential retention of some ½<111> loop variants in non-<100> grains was thought to promote elastic interactions of same-Burgers-vector loops and thereby string structure formation [7]. The effects of temperature and alloying can be attributed to the mobility of dislocation loops, which can be enhanced by the increase of temperature, and reduced by the addition of alloying elements.

Since the formation of complex string structure is not favoured in <100> grains at RT, the observations of complex string structures in this study may in part be ascribed to localized heating in the region of dense collisions [41]. The formation of string structure involves two different mechanisms: loop-loop interaction and loop-string interaction, that have been observed by TEM [16,20]. Loop-loop interaction refers to the elastic interaction of neighbouring loops. There are two different scenarios for loop-loop interaction. In the first scenario, two loops, if close enough, can directly coalesce to form a larger dislocation loop. In this process, the interacting loops can either have the same Burgers vector or different Burgers vectors [16,20]. In the second scenario, two or more loops move cooperatively until the minimum elastic interaction energy is achieved at the 'string-configuration'. This scenario can also be understood as the mutual pinning of small dislocation loops at short distances [26,74]. In this process, the loops involved usually have the same Burgers vector. However, it is not excluded that some additional clusters or small loops can be trapped by the pinned structure by changing their Burgers vector [74]. The mutual pinning of dislocation loops is believed to be the formation mechanism of the string configurations seen by ECCI in the 0.1 dpa sample in Fig. 2(c) (pointed out by arrows) and Fig. 10(b).

However, most loops in the 0.1 dpa sample are not well aligned into string configurations. Instead, they tend to form clusters of loops. This is in good agreement with TEM observation on a FIB lift-out from 0.1 dpa 20 MeV self-ion irradiated tungsten [41]. When viewing the implanted layer cross-section, the formation of string structures showed a strong depth dependence. Near the surface, loops have a lower tendency to organise into string configurations [41]. This is probably because the damage level at the surface is lower than the peak damage dose. The evolution of the defect structure observed ECCI (Fig. 2) suggests that clustering of dislocation loops marks the transition from a random distribution of loops at doses below cascade overlap, to string structures at higher doses. This stage can be understood as the onset of heterogeneity of defect structure. The clusters have a characteristic length scale of ~ 300 nm (Fig. 4), which indicates that loop-loop interactions at 0.1 dpa transition from short range to a long-range many-body problem. If loops of a specific Burgers vector dominate within this range, the interaction may lead to the string configuration of the loops. Otherwise, it results in a clustering configuration. More loops configurated into the form of string structure with the increase of dose. Therefore, it is expected that loops within a cluster may change their Burger vector either by being absorbed by larger loops [16] or by the elastic strain field of stable sub-cluster of loops [74]. Also, it can not be excluded that the overlap of a new collision cascade the pre-existing loop can produce a new loop with Burgers vector which tends to reduce the overall elastic energy.

### 4.3. Fluctuation in lattice strain and rotations

The striking finding of the HR-EBSD strain measurement is a transition from a lack of strain fluctuation at 0.01 to long-range strain fluctuations at doses of 0.1 dpa and greater. The magnitude of fluctuation increases with damage dose until 0.32 dpa, where saturation is achieved. The characteristic length scale associated with these fluctuations above 0.32 dpa is ~500 nm. The evolution of these strain fluctuations is in line with the evolution of the defect structure. At 0.01 dpa, dislocation loops are distributed uniformly and randomly, and there are no visible long-range strain fluctuations. As dose increases to 0.1 dpa, dislocation loops arrange into clusters or chain structures, and strain fluctuations begin to form (Fig. 10(b)). At 1 dpa, more string structures form, and the individual loops that constitute the strings become difficult to resolve (Fig. 10(c)). In the strain maps, more obvious fluctuations are observed. The parallel evolution of the defect structure and strain fluctuations indicates that these two aspects are linked: They are both the results of the long-range many-body elastic interactions of dislocation loops during the continuing generation of new collision cascade damage. The driving force for the development of strain fluctuations is presumably the minimization of the elastic strain energy of existing dislocation loops and newly generated dislocation loops. The saturation of the strain fluctuations may indicate that a stable elastic energy state can be achieved above a certain dose (0.32 dpa in this study). That is, beyond 0.32 dpa, the overall elastic energy stored in defects no longer increases significantly.

To quantitatively study the correlation between the defect structures and the strain fluctuations will require dislocation dynamics simulation that is beyond the scope of this paper. The parallel evolution of strain fluctuations and long-range spatial ordering of dislocation loops suggests that strain fluctuation could be used as a proxy to assess the extent of spatial ordering of defects structure. As IIDs in bcc material [26] and hcp materials [8,75,76] tend to form ordered structures, HR-EBSD presents a convenient tool for the non-destructive characterization of defect evolution in these materials.

## 5. Conclusion

In this paper, we demonstrate the prototypical non-destructive characterization of defect evolution in 20 MeV self-ion irradiated tungsten using SEM-based techniques. Electron channelling contrast imaging is used to characterize the microstructural evolution of the defects. Cross-correlation-based HR-EBSD makes it possible to probe the evolution of lattice strain and rotation fields at the sample surface. Together, these techniques open up interesting new opportunities for the characterisation of irradiation damage. Several conclusions can be drawn from the experimental observations.

1) ECCI reveals the evolution of defect structures from 0.01 dpa to 1 dpa in the following sequence: spatially random distribution of dot-like dislocation loops → clustering of dot-like dislocation loops with increased size → a mixture of string structures and dotty loops. The characteristic lengthscale of clustering of defects is 300 nm and 500 nm at 0.1 dpa and 1 dpa, respectively.

2) Observations from ECCI are qualitatively in good agreement with previous TEM studies. ECCI has the potential to enable a more rapid throughput characterisation for the study of defect structure evolution as a function of irradiation damage. It also avoids the ambiguities introduced by FIB-based sample preparation.

3) Using HR-EBSD the nano-scale lattice strains and rotations introduced by irradiation defects can be probed. Our results show that there is little strain fluctuation at 0.01 dpa though the defect density is close to the saturation value. Beyond 0.1 dpa, strain fluctuations with a wavelength on the order of 500 nm appear, which agrees well with the characteristic lengthscale of clustering of defects. This shows that there is a transition from long-range randomness to long-range spatial ordering of dislocation loops with increasing damage dose.

4) The saturation of strain fluctuations and the GND density above 0.32 dpa suggests that the elastic energy stored in irradiation-induced defects will eventually approach a stable value. After the GND saturation dose, the GND density measured from HREBSD can be used as a very useful parameter to be included in mechanical models that seek to capture irradiation-induced changes in mechanical properties.

**Acknowledgements**

This work was funded by Leverhulme Trust Research Project Grant RPG-2016-190. The authors acknowledge the use of characterisation facilities within the David Cockayne Centre for Electron Microscopy, Department of Materials, University of Oxford. Junliang Liu is grateful for support from EPSRC grant (EP/P001645/1). The Zeiss Crossbeam FIB/SEM used in this work was supported by EPSRC through the Strategic Equipment Fund, grant #EP/N010868/1.